# Comment on "Superconductivity in electron-doped layered TiNCl with variable interlayer coupling"

Dale R. Harshman
*Physikon Research Corporation, Lynden, Washington 98264, USA
and Department of Physics, University of Notre Dame, Notre Dame, Indiana 46556, USA*

Anthony T. Fiory
*Department of Physics, New Jersey Institute of Technology, Newark, New Jersey 07102, USA*

## Abstract

In their article, Zhang *et al*. [Phys. Rev. B **86**, 024516 (2012)] present a remarkable result for $A_x(S)_y$TiNCl compounds (α-phase TiNCl partially intercalated with alkali $A$ and optionally co-intercalated molecular species $S$), finding the superconducting transition temperature $T_C$ scales with $d^{-1}$, where the spacing $d$ between TiNCl layered structures depends on intercalant thickness. Recognizing that this behavior indicates interlayer coupling, Zhang *et al*. cite, among other papers, the interlayer Coulombic pairing mechanism picture [Harshman *et al*., J. Phys.: Condens. Matter **23**, 295701 (2011)]. This Comment shows that superconductivity occurs by interactions between the chlorine layers of the TiNCl structure and the layers containing $A_x$, wherein the transverse $A_x$-Cl separation distance $\zeta$ is smaller than $d$. In the absence of pair-breaking interactions, the optimal transition temperature is modeled by $T_{C0} \propto (\sigma/A)^{1/2}\zeta^{-1}$, where $\sigma/A$ is the fractional charge per area per formula unit. Particularly noteworthy are the rather marginally-metallic trends in resistivities of $A_x(S)_y$TiNCl, indicating high scattering rates, which are expected to partially originate from remote Coulomb scattering (RCS) from the $A_x$ ions. By modeling a small fraction of the RCS as inducing pair-breaking, taken to cut off exponentially with $\zeta$, observations of $T_C < T_{C0}$ are quantitatively described for compounds with $\zeta < 4$ Å, and $T_C \approx T_{C0}$ for Na$_{0.16}(S)_y$TiNCl with propylene carbonate and butylene carbonate co-intercalants for which $\zeta > 7$ Å. Since a spatially separated alkali-ion layer is not formed in Li$_{0.13}$TiNCl, the observed $T_C$ of 5.9 K is attributed to an intergrowth phase related to TiN ($T_C = 5.6$ K).

## I. INTRODUCTION

The layered compound TiNCl belongs to the group 4 metal nitride-halides of composition $M$N$X$ with $M$ = Ti, Zr, or Hf and $X$ = Cl, Br, or I, forming $X$-($M$N)$_2$-$X$ layered structures with van der Waals bonding between twin halide layers ($X$-$X$). Pristine $M$N$X$ compounds of principal crystal structures termed α and β forms are considered to be band insulators or wide-gap semiconductors.[1-4] Intercalation doping of a cationic element ($A_x$) between the halide layers, including co-intercalated molecular solvent species ($S_y$), induces superconductivity in $A_x(S)_yMNX$ systems. Probably the most intriguing behavior observed is the strong correlation of $T_C$ with the basal-plane separation $d$, defined as the periodicity distance along the $c$-axis between $[MNX]_2$ blocks.[5] The subject paper by Zhang et al.[6] on $A_x(S)_y$TiNCl compares eight compounds with relatively dilute doping x of an alkali element ($A$ = Li, Na, K, and Rb) including four with co-intercalated molecules ($S$ = THF, PC, and BC, denoting tetrahydrofuran, polypropylene carbonate, and butylene carbonate, respectively). The authors of Ref. 6 show that, with the exception of $Li_{0.13}$TiNCl, there exists a linear dependence between $T_C$ and $1/d$, which is accepted as strong evidence of interlayer coupling, and suggest the relevance of our study of transition temperatures in high-$T_C$ superconductors.[7] Data for $T_C$ and $d$, as tabulated and read from Fig. 3 of Ref. 6, are listed in Table I.

## II. $T_C$ MODEL

For $A_x(S)_y$TiNCl, the interlayer interaction model follows from the identification of two types of layered charge reservoir structures,[7] in which the $[TiNCl]_2$ structure is proposed for type I and the intercalant layer $A_x(S)_y$ structure is proposed for type II, the latter including $A_x$-intercalation at y = 0. The type I reservoir hosts and sustains the superconducting current, whereas the type II reservoir provides the mediation for the superconductive pairing interaction. In this model, the coupling occurs between adjacent ionic layers and thus involves the outer chlorine layers in $[TiNCl]_2$ and the alkali $A_x$ in the intercalant layers. The optimal transition temperature $T_{C0}$ occurs on the formation of participating charges in the two reservoirs for x at optimal doping.

These compounds present a unique situation among high-$T_C$ materials in which the mediating layer adjacent to the superconducting condensate is incomplete, containing a high density of vacancies. Even without disorder in the $[TiNCl]_2$ bilayer structures,[8] the close proximity of disordered $A$ cations can induce remote Coulomb scattering (RCS) analogous to the depression of carrier mobility in two-dimensional systems by remote fixed charges.[9] Evidence for the presence of significant scattering are large resistivities at $T_C$ and broad resistive transitions $\Delta T_C$.[6] It is found herein that pair breaking via RCS limits the superconductivity from achieving an optimal state in $A_x(S)_y$TiNCl compounds with $d < 20$ Å, causing $T_C < T_{C0}$.

Given this interaction model and structure, it is possible to explain all eight of the data points given in Fig. 3 of Ref. 6, including that of $Li_{0.13}$TiNCl which does not follow the linear trend. In applying the model, one proceeds under the caveat that the results for $T_C$ are obtained for compositions at or near optimal.

### A. Optimal $T_{C0}$

High-$T_C$ superconductivity in this model occurs in layered structures forming adjacent type-I and type-II charge reservoir layers containing the superconducting and mediating charges, respectively, repeating alternately along the transverse axis. The superconducting transition temperature depends on the spatially indirect Coulomb interaction across the transverse distance $\zeta$ between the two charge reservoirs, measured between the outer chlorines in the type-I $[TiNCl]_2$ layer and the locus of the cations $A_x$ in the neighboring type II intercalation layer, assuming co-intercalant $(S)_y$ is uncharged. The layered structure of $A_x(S)_y$TiNCl is characterized by a thickness $d_2$ of the $[TiNCl]_2$ layers, the transverse spacing $d$ between them, and an intercalant thickness $d - d_2$.[1] Assuming that the mean cation $A_x$ locus is at the intercalant-layer midplane, the interaction distance is $\zeta = (d - d_2)/2$. Since $d_2$ is approximately the same as for pristine $\alpha$-TiNCl,[1] the observed functional dependence of $T_C$ on $d$ is expected to correlate with an analogous dependence on $\zeta$. However, one notes that the interlayer interaction length is the shorter distance $\zeta$, rather than the spacing $d$. Structural and superconductivity data are presented in Table I, listing directly measured values of $d_2$ where available. The Coulomb energy $e^2/\zeta$ lies within 1.8 – 8.7 eV.

TABLE I. Structural and electronic parameters for $A_x(S)_y$TiNCl.

| Compound | $T_C$ (K) | $d$ (Å) | $d_2$ (Å) | $A$ (Å$^2$) | $\zeta$ (Å) | $T_{C0}$ (K) | $\alpha$ (meV) | $T_C^{calc.}$ (K) |
|---|---|---|---|---|---|---|---|---|
| Na$_{0.16}$TiNCl | 18.0 | 8.442 | 5.150 | 13.1564 | 1.6460 | 29.55 | 1.175 | 18.20 |
| Na$_{0.16}$(THF)$_y$TiNCl | 10.2 | 13.105 | 5.183 | 12.9753 | 3.9610 | 12.36 | 0.230 | 10.35 |
| Na$_{0.16}$(PC)$_y$TiNCl | 7.4 / 6.3 | 20.53 | 5.183 | 13.0331 | 7.6735 | 6.37 | 0 | 6.24 |
| Na$_{0.16}$(BC)$_y$TiNCl | 6.9 | 20.7435 | 5.183 | 13.0331 | 7.7803 | 6.28 | 0 | 6.16 |
| K$_{0.17}$TiNCl | 17.0 | 8.77884 | 5.182 | 13.3720 | 1.7984 | 27.65 | 1.086 | 16.84 |
| Rb$_{0.24}$TiNCl | 16.0 | 9.21038 | 5.000 | 13.2830 | 2.1052 | 28.16 | 1.225 | 15.81 |
| Li$_{0.13}$(THF)$_y$TiNCl | 9.5 | 13.0012 | 5.183 | 13.1277 | 3.9091 | 11.23 | 0.184 | 9.53 |
| Li$_{0.13}$TiNCl | 5.9 | 7.82451 | 5.133 | 13.1277 | | | | |

The doping-dependent optimization behavior, as generally exhibited by high-$T_C$ compounds, suggests that optimization corresponds to equilibrium between the two reservoirs. The two-dimensional density of interaction charges is given in the model as $\sigma\eta/A$, where $\sigma$ is the participating fractional charge in the type-I reservoir, determined per formula unit by doping as discussed below, $\eta$ is the number of charge carrying layers in the type-II reservoir and is given by $\eta = 1$ for $A_x(S)_y$, and $A$ is the crystal basal-plane area per formula unit. Adopting this approach, it has been shown that the optimal transition temperature $T_{C0}$ is given by the algebraic expression $k_B^{-1}\,\beta\,\zeta^{-1}\,(\sigma\eta/A)^{1/2}$, where $\beta = 0.1075(3)$ eV Å$^2$ is the universal constant determined previously by fitting experimental data.[7] The modeled $T_{C0}$ is the upper limit on the experimentally observed transition temperature, given $T_C < T_{C0}$ for non-optimal materials.[10]

Particularly important to this model is the concept of the participating charge that is defined for optimal materials as the difference between the dopant charge stoichiometry and the minimum stoichiometric value required for superconductivity. An example is $x = 0.163$ taken relative to $x_0 = 0$ in La$_{2-x}$Sr$_x$CuO$_{4-\delta}$. Doping is by direct cationic or anionic substitution in one or both reservoirs.[11,12] For $A_x(S)_y$TiNCl, doping occurs only in the intercalation layer via $A_x$, such that $\sigma$ is determined according to the simplified relation,

$$\sigma = \gamma \,|\, v\,(x - x_0)\,|, \qquad (1)$$

where $v$ is the valence and $x$ is the optimal content of the cation dopant species in the type II $A_x(S)_y$ reservoir; $x_0$ is the threshold value of $x$ for superconductivity; here, $v = 1$ for alkali-ion doping and $x_0 = 0$ is inferred from Refs. 1 and 6. The factor $\gamma$ derives from the allocation of the dopant by considering a given compound's structure. Following the procedure generally applied to high-$T_C$ superconductors, the charge introduced by the dopant is shared equally between the two charge reservoirs. Additionally, the methodology requires the doped charge to be distributed pair-wise between the charge-carrying layer types within each of the charge reservoirs. Assuming the co-intercalant contributes no doping charge, determination of $\gamma$ for $A_x(S)_y$TiNCl is comparable to that of (Ba$_{0.6}$K$_{0.4}$)Fe$_2$As$_2$,[7] for which a structural analogy was previously noted.[13] Sharing the charge equally between the two reservoirs contributes a factor 1/2 to $\gamma$. Sharing between the Cl layer and the double-TiN layered structure and then to the two TiN layers contributes two factors of 1/2 to $\gamma$. Hence, $\gamma = (1/2)(1/2)(1/2) = 1/8$, yielding $\sigma$ generally smaller than $x$.

## B. Pair-breaking Scattering

Although intercalation doping induces superconductivity in $A_x(S)_y$TiNCl, $\Delta T_C$ is broad and resistivity near $T_C$ is high and semiconductor-like, e.g., data for $Na_{0.16}$TiNCl show $T_C = 18.0$ K, $\Delta T_C \approx 5$ K relative to the transition midpoint, and resistivity $\rho(T_C^+) \approx 0.27$ $\Omega$ cm exhibiting an upturn as $\rho^{-1}Td\rho/dT \approx -0.3$ just above $T_C$.[6] Recognizing that the high measured resistivities have been attributed to the polycrystalline morphology of the samples under study,[6,14] these are also signatures of high electron scattering rates $\tau^{-1}$. A likely scattering mechanism is found by drawing analogy to modulation doping of semiconductor quantum wells[15] or RCS of carriers in a semiconducting inversion layer from fixed charges located outside the layer.[9] The form factor for the scattering process follows from the indirect Coulomb potential $v(q) \propto \exp(-qz)$, where $q$ is the scattering wavevector and $z$ is the transverse distance between the conducting plane and the location of the Coulomb scattering center. Since the damping factor is obtained from an integration over potential fluctuations scaling with $|v(q)|^2$,[16] one expects in application of RCS to $A_x(S)_y$TiNCl that $\hbar\tau^{-1}$ attenuates exponentially with the product of a characteristic value for $q$ and $z$ given by $\zeta$ or $d/2$.

Under conditions of strong scattering, particularly in the limit of small $\zeta$ where RCS would be strongest, it is possible that some fraction of the scattering contributing to $\hbar\tau^{-1}$ also induces pair-breaking scattering in the superconductor. Analogous pair-breaking effects in the cuprates originate from magnetic impurity scattering[17] and disorder associated with non-optimization[10,18] These pair-breaking effects, causing the observed depression of $T_C$ below $T_{C0}$, are distinguished from weak scattering phenomena, which are less likely to affect $T_C$.[19] The following expression describes the pair-breaking affect on $T_C$,[20] which has been applied to treat disorder in thin films[21] and the cuprates:[10,18]

$$\ln(T_{C0}/T_C) = \psi(\tfrac{1}{2} + \alpha/2\pi k_B T_C) - \psi(\tfrac{1}{2}). \quad (2)$$

Here, $\psi$ is the digamma function, $T_C$ is the experimentally measured transition temperature, $T_{C0}$ is the optimal transition temperature calculated by assuming no pair breaking (Sec. II.A), and $\alpha$ is the pair-breaking parameter. Thus, for a given compound with measured $T_C$ and calculated $T_{C0}$, one obtains the associated $\alpha$ from Eq. (2).

Where the ionized intercalant induces RCS, pair-breaking is modeled by scaling $\alpha$ to the valence v and content x of species $A_x$ and an exponential attenuation factor wherein the transverse distance is taken as $\zeta$:

$$\alpha = a_1 vx \exp(-k_1\zeta). \quad (3)$$

Equation (3) is expressed in terms of empirical parameters, the coefficient $a_1$ and attenuation rate $k_1$, which incorporate by approximation dependencies on scattering wave vector and screening as well as the finite thicknesses of the [TiNCl]$_2$ and $A_x(S)_y$ layers. In this model the pair-breaking rate, given by $2\hbar^{-1}\alpha$,[20] is expected to be small compared to the total scattering rate $\tau^{-1}$ associated with electrical transport.

In principle, screening dominates the strength of Coulomb scattering.[9] Because the bulk of the superconducting current flows in the [TiN]$_2$ substructures, one would expect $T_C$ to vary significantly between the $\alpha$ and $\beta$ forms of $A_x(S)_yMNX$, and be modulated by the screening effects of the species $M$, $X$, and $A$, which increase with their atomic Z. The comparatively weaker cation screening available in the $\alpha$-TiNCl compounds is expected to produce greater RCS-related suppression in $T_C$ when compared to compounds based on $\beta$-ZrNCl or $\beta$-HfNCl, owing to the comparatively larger Z of Zr or Hf and larger $d_2$ of the $\beta$ form.[1] This may account for results reported for Li$_x$ZrNCl, where mobilities and mean-free-paths derived from $H_{C2}(T)$ data indicate minimal disorder scattering from the Li intercalant.[15]

## III. EXPERIMENTAL $T_C$

The starting point for understanding experimental results for $T_C$ in $A_x(S)_y$TiNCl is the optimal transition temperature $T_{C0}$ calculated for the unique optimal doping x,

$$T_{C0} = k_B^{-1} \beta \zeta^{-1} [0.125\, v(x - x_0)/A]^{1/2}. \quad (4)$$

Using this model for $T_{C0}$ in conjunction with the pair-breaking expression of Eq. (3), one may understand the variation in $T_C$ with $d$ for $A_x(S)_y\text{TiNCl}$.[6] Since $\zeta$ depends functionally on $d$, correlations between $T_C$ and $d$ are thus possible, owing to the approximately constant $d_2$ (see Table I).

Zhang et al.[6] found that the correlation of $T_C$ with $1/d$ breaks down for $\text{Li}_{0.13}\text{TiNCl}$, indicating that the length $d$ is, perhaps, not the relevant length parameter involved. As $d$ is not directly associated with a specific pair of interacting layers, this is not surprising. The key difference between $d$ and $\zeta$ is that the former is always defined and non zero, whereas the latter is unrealized in the absence of two physically-separated and adjacent interacting layers. This subtle, but key, difference explains why $\text{Li}_{0.13}\text{TiNCl}$ does not behave in the same manner as the other seven compounds, and provides strong support for the interlayer Coulomb interaction model described in Sec. II. Since the Li cations occupy sites between the Cl anions,[6,22] a spatially separated intercalation layer is not formed; hence, $\text{Li}_{0.13}\text{TiNCl}$ does not possess the requisite two-layer interaction structure. The measured $T_C$ is therefore hypothesized to reflect the BCS superconductivity of an intergrowth phase or inclusions related to TiN which has $T_C = 5.6$ K. This structural distinction also explains the absence of superconductivity in $\text{H}_x\text{ZrNCl}$,[23] where in this case the H impurities occupy the 6c site between the Zr-N and the Cl ions and dope the type-I reservoir; since the type II reservoir is absent, high-$T_C$ superconductivity does not occur. Non-superconducting Li-doped α-phase HfNBr appears to be a similar case, in which localized spin paramagnetism is formed.[24] Additionally, Zhang et al.[6] compare their results with earlier studies of intercalated Bi-based cuprates,[25] in which intercalation of charge-neutral molecules between the double BiO layers leaves $T_C$ unchanged. Since $\zeta$ for the Bi-based cuprates is defined as the distance between adjacent SrO and $\text{CuO}_2$ layers,[7] which does not change upon intercalation, this behavior is expected and confirms that $\zeta$, not $d$, is the length which governs $T_{C0}$ in high-$T_C$ superconductors.

As evident from upturns in resistivity just above $T_C$, there exist large background scattering effects in the $A_x(S)_yMNX$ systems. High scattering rates can sometimes result in pair breaking via RCS interactions, degrading the superconducting state and forcing $T_C$ below $T_{C0}$. In particular, the α-TiNCl-based materials, exhibiting comparatively higher resistivities just above $T_C$,[6] relative to those based on the β forms of ZrNCl[26,27] and HfNCl,[5,14] are certainly good candidates. Following the logic set forth in Sec. II, the task becomes one of identifying and quantitatively extracting the pair-breaking component. To accomplish this, one first calculates $T_{C0}$ assuming optimization; materials free of pair breaking and possessing the optimal cation doping necessarily exhibit $T_C = T_{C0}$ (within uncertainties). The suppression of $T_C$ below $T_{C0}$ evident in the remaining compounds can then be attributed to RCS-induced pair breaking or other pair-breaking phenomena.

Absent structural refinement data, the value of $d_2$ assumed from a related material or the host (see Fig. 4 and table 4 of Ref. 1) is used in determining $\zeta$. Values for $\zeta$ and $T_{C0}$ from Eq. (4) are shown in Table I for the seven high-$T_C$ compounds of Ref. 6. As can be seen, only the two compounds with the largest $\zeta$, $\text{Na}_{0.16}(\text{BC})_y\text{TiNCl}$ and $\text{Na}_{0.16}(\text{PC})_y\text{TiNCl}$, can be considered optimal, having $T_C \approx T_{C0}$ (Table I includes $T_C = 6.3$ K obtained by extrapolating $H_{C2}(T)$ for $\text{Na}_{0.16}(\text{PC})_y\text{TiNCl}$ in Fig. 2(b) of Ref. 6), whereas the others show progressively larger deviation from optimal behavior with decreasing $\zeta$. These deviations of $T_C < T_{C0}$, interpreted in terms of pair-breaking, determine finite values of the pair-breaking parameter α as solutions of Eq. (2). The two compounds with $T_C \approx T_{C0}$ are taken to have α = 0. The resulting values for α are listed in Table I. One finds that $A_x\text{TiNCl}$ without co-intercalation molecules exhibit the highest α values, 1.09–1.35 meV, as expected for minimum $\zeta$. The pair-breaking rate associated with 2α, which is less than 2.7 meV, is a very small component of the total scattering rate contained in $\hbar\tau^{-1}$. This can be ascertained, for example, by considering the damping factors $\hbar\tau^{-1} > 0.2$ eV indicated optically for Li$_x$ZrNCl,[28] and noting that transport measurements indicate higher resistivities for α-TiNCl-based compounds (e.g., $\rho(T_C^+) \approx 0.27$ Ωcm for Na$_{0.16}$TiNCl[6]) when compared to β-ZrNCl-based compounds (e.g., $\rho(T_C^+) \approx 6.2$ mΩcm for Li$_{0.08}$ZrNCl[15]). Hence, these results are consistent with having $2\alpha \ll \hbar\tau^{-1}$.

In view of Eq. (3) based on the RCS model, the pair-breaking parameters for the seven $A_x(S)_y$TiNCl compounds were scaled with doping in the form α/x (v = 1 for the alkali ions) and plotted in Fig. 1 to show dependence on ζ. The curve is a fit to the function $a_1\exp(-k_1\zeta)$, with $a_1 = 23.9 \pm 1.0$ meV and $k_1 = 0.727 \pm 0.023$ Å$^{-1}$, which displays remarkable representation of the data; the root-mean-square (rms) deviation between the α/x data and the corresponding function is 0.10 meV. In the limit where 2ζ approaches the van der Waals gap of pristine TiNCl (2.618 Å)[1], the hypothetical maximum α of (9.3 ± 0.4 meV) vx is also small compared to reasonable estimates of $\hbar\tau^{-1}$. The attenuation factor provides an estimate of the characteristic pair-breaking scattering wave vector as $\langle q \rangle = k_1/2 \sim 0.42$ $\pi A^{-1/2}$, suggesting large-angle scattering dominates ($\pi/A^{1/2} \approx 0.868$ Å$^{-1}$ from Table I). Modeling the distance $z$ as $d/2$ in place of ζ in Eq. (3), yields $a_1 = 160 \pm 40$ meV, $k_1 = 0.736 \pm 0.042$ Å$^{-1}$, and 0.18 meV rms deviation; the larger error obtained with $z = d/2$ correlates with the small variations in $d_2$.

Evaluation of the linear trend between $T_C$ and $1/d$ noted in Ref. 6 is readily obtained by fitting the function $T_C = s/d$ to the data for $A_x(S)_y$TiNCl (excluding Li$_{0.13}$TiNCl), yielding $s = 145 \pm 4$ Å K and rms deviation of 0.87 K in $T_C$. In comparison, the calculated transition temperature $T_C^{\text{calc.}}$, as determined by Eq. (2) with $T_{C0}$ from Eq. (4) and α from Eq. (3), yields only 0.54-K rms deviation from measured $T_C$, indicating significant improvement for the model-based analysis over the heuristic scaling with $1/d$. The results for $T_C^{\text{calc.}}$ are given in Table I. Note that without RCS-related pair breaking, $T_C = T_{C0}$ and would approach 30 K for $A_x$TiNCl without co-intercalant molecules.

## IV. CONCLUSION

The very fine work of Zhang et al.[6] on α-form polymorphs $A_x(S)_y$TiNCl is interpreted from the perspective of an interlayer Coulombic interaction model,[7] identifying the superconducting [TiNCl] and mediating [$A_x(S)_y$] charge reservoirs, the relevant interaction distance ζ measured between cations $A_x$ and Cl (different from the basal-plane separation $d$), and optimal transition temperature $T_{C0}$. Recognizing the presence of strong scattering from transport data, it is postulated that the transition temperatures could be suppressed due to pair breaking arising from the proximity of the superconducting layer to the dilute, disordered charges of the intercalation layer. By adapting a pair-breaking model based on RCS, it is shown that the maximum attainable $T_C$ ($\leq T_{C0}$) is determined by a unique pair-breaking function α, which falls off exponentially with increasing ζ. Not unexpectedly, the two compounds with the largest interaction distances, Na$_{0.16}$(PC)$_y$TiNCl and Na$_{0.16}$(BC)$_y$TiNCl (ζ = 7.6738 Å and 7.7803 Å, respectively), are found to be optimal with $T_C \approx T_{C0}$, whereas for the others (apart from Li$_{0.13}$TiNCl) RCS pair breaking is more dominant, owing to smaller ζ, yielding $T_C < T_{C0}$.

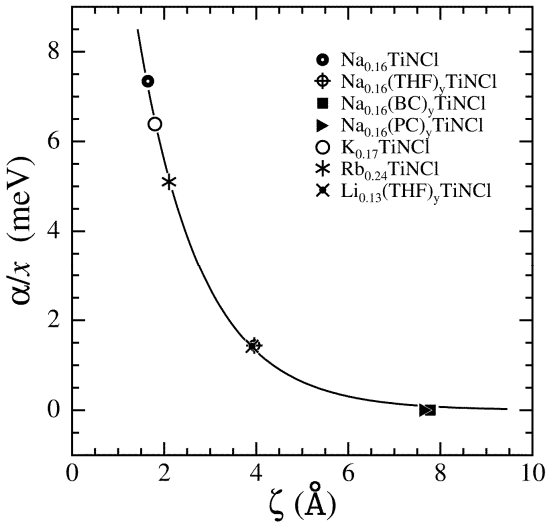

FIG. 1. Reduced pair-breaking parameter α/x plotted against interaction distance ζ for $A_x(S)_y$TiNCl. The curve is the fitted function of Eq. (3).

With the understanding that the important length governing $T_C$ is $\zeta$ and not $d$, the anomalous behavior of Li$_x$TiNCl[6] and H$_x$ZrNCl[23] is attributed to the location of the dopants in the [MNCl]$_2$ layers, such that the physically separated mediating layer for high-$T_C$ superconductivity is not formed and $\zeta$ is unrealized. This result suggests that the superconductivity observed in Li$_x$TiNCl is related to the BCS superconductivity of TiN ($T_C$ = 5.6 K).[6,13,22]

## ACKNOWLEDGMENTS

We are grateful for the support of the Physikon Research Corporation (Project No. PL-206) and the New Jersey Institute of Technology. We also thank Dr. S. Zhang and Professor R. F. Marzke for providing helpful and important information. This paper is published as a comment[29] and is accompanied by a reply.[30]